\documentclass[%twocolumn,%showpacs,preprintnumbers,amsmath,%
amssymb,aps,
prd,nofootinbib,floatfix,12pt,superscriptaddress,%preprint,
]{revtex4}%
\usepackage[T1]{fontenc}
\usepackage[utf8]{inputenc}
\usepackage{textcomp}
\usepackage{amssymb}
\usepackage{amsmath}
\usepackage{epsfig}
\usepackage{graphicx}
\usepackage{esint}
\usepackage{comment}
\usepackage{xcolor}
\usepackage{color}
\PassOptionsToPackage{normalem}{ulem}
\usepackage{ulem}
\usepackage{slashed}
\usepackage{caption}
\usepackage{amsmath,amssymb,epsf}
\usepackage{graphicx,wrapfig,color}
\usepackage{subfig}
\usepackage{setspace}
\usepackage{makecell}
\usepackage{tabularx}
\usepackage{booktabs}
\usepackage[tablename=Table]{caption}
\usepackage{float}
\usepackage{placeins}
\usepackage[unicode=true,pdfusetitle,
 bookmarks=true,bookmarksnumbered=false,bookmarksopen=false,
 breaklinks=false,pdfborder={0 0 0},backref=false,colorlinks=true,citecolor=blue,urlcolor=violet 
]{hyperref}

\usepackage{xcolor}
\usepackage{footnote}
\newcommand{\orcid}[1]{\href{https://orcid.org/#1}{\includegraphics[width=10pt]{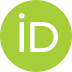}}}
\newcommand{\be}{\begin{equation}}
\newcommand{\ee}{\end{equation}}
\newcommand{\bea}{\begin{eqnarray}}
\newcommand{\eea}{\end{eqnarray}}

\newcommand{\ba}{\begin{array}}
\newcommand{\ea}{\end{array}}
\newcommand{\bd}{\begin{displaymath}}
\newcommand{\ed}{\end{displaymath}}

\def\gev{{\rm \,Ge\kern-0.125em V}}
\def\tev{{\rm \,Te\kern-0.125em V}}

\def\th13 {\theta_{13}}

\begin{document}

\title{ Production of Primordial Black Holes via Single Field Inflation and Observational Constraints }

\author{Mayukh R. Gangopadhyay\orcid{0000-0002-1466-8525}}
\email{mayukh_ccsp@sgtuniversity.org}
\affiliation{Centre for Cosmology and Science Popularization, SGT University, Gurugram, Haryana-122505, India}

\author{Jayesh C. Jain}
\email{419ph2110@nitrkl.ac.in}
\affiliation{National Institute of Technology Rourkela, Odisha 769001, India.}

\author{Devanshu Sharma}
\email{i17ph001@phy.svnit.ac.in}
\affiliation{Sardar Vallabhbhai National Institute of Technology Surat, Gujarat 395007, India.}

\author{Yogesh\orcid{0000-0002-7638-3082}}
\email{yogesh.ccsp@sgtuniversity.org,yogesh@ctp-jamia.res.in}
\affiliation{Centre for Cosmology and Science Popularization, SGT University, Gurugram, Haryana-122505, India}
\affiliation{Centre for Theoretical Physics, Jamia Millia Islamia, New Delhi-110025, India.}

\iffalse
\author{$^1$Mayukh~R.~Gangopadhyay, $^{2}$Jayesh C. Jain, $^{3}$Devanshu Sharma, $^4$Yogesh}
\affiliation{
\it$^{1,4}$Centre for Cosmology and Science Popularization, SGT University, Gurugram, Haryana-122505, India.\\
\it$^{4}$Centre For Theoretical Physics,Jamia Millia Islamia, New Delhi-110025, India.\\
\it$^2$ National Institute of Technology Rourkela, Odisha 769001, India.\\
 \it$^3$Sardar Vallabhbhai National Institute of Technology Surat, Gujarat 395007, India.\\}
   
\email[]{$^1$mayukh@ctp-jamia.res.in\\$^2$419ph2110@nitrkl.ac.in\\$^3$i17ph001@phy.svnit.ac.in\\$^4$yogesh@ctp-jamia.res.in}

\fi
%\date{\today}
%%%%%%%%%%%%%%%%%%5
\begin{abstract}
In a class of  single field models of inflation, the idea of Primordial Black holes(PBHs) production is studied. In this case, the dynamics on small cosmological scales differs significantly  from that of the large scales probed by the observations of cosmic microwave background(CMB). This difference becomes a virtue in producing correct physical ambiance for the seeds required to produce PBHs. Thus, once the perturbed scales renter the horizon of our Universe during the later epochs of radiation domination and subsequent matter domination, these seeds collapses to produce PBHs. We have shown, in this class of model, depending on the model parameters and the class defining set parameters, one can have PBHs formed for a vast mass ranges from $10^{-18}$ to $10^{-6}$ solar mass(\(\textup{M}_\odot\)). We have also shown, for a particular class of model, the total dark matter density today can be attributed to the PBHs density. The vast range of the mass depending on the class parameter,  gives ample opportunity to study enriched phenomenological implications associated with this model to probe the nascent Universe dynamics.
\end{abstract}
\maketitle

\vspace{1in}

\iffalse
\tableofcontents
\fi
\baselineskip=15.4pt

\vspace{1cm}
%%%%%%%%%%%%%%%%%%%%%
\section{Introduction}
%%%%%%%%%%%%%
%%%%%%%%%%%%%
The inflationary scenario is a remarkable and compelling paradigm that stands as one of the fundamental pillars of modern cosmology. By cosmological inflation, one refers to a brief period of accelerated expansion during the first moment of the Universe right before the radiation dominated epoch. Originally, inflation was invoked to overcome some of the problems associated with the hot big bang model such as horizon and flatness problem{\cite{Guth,Linde,Baumann,liddle,albrecht,turner1,lyth,ijjas,Weinberg,Dodelson,Mukhanov}}. However, currently, inflation proves to be the most dominant paradigm for the origin of cosmological perturbations in the early Universe. Cosmological inflation can also produce the seeds of the Primordial Black Holes(PBHs) \cite{hawking}. These seeds can later produce the PBHs in the radiation dominated or even matter-dominated era. Thus formation and evolution of PBHs can be a great way to probe early time Cosmology \cite{Carr:2016drx}. The possibility of the PBHs was first proposed by Zeldovich and Novikov \cite{Zeldovich}.\\*
 In 1974, Prof. Stephen Hawking made a remarkable proposal that black holes can emit thermal radiation due to the quantum effects (Hawking radiation). According to his proposal, black holes with smaller masses would evaporate at a faster rate \cite{Hawking,Don}. Thus, primordial black holes (PBHs) are of particular interest as PBHs were formed in the early Universe and effectively can have mass as low as Planck mass unlike the black holes formed by the collapse of the stellar objects which has to have a minimum mass of 1.4 times the \(\textup{M}_\odot\) (Chandrasekhar limit). Also PBHs with larger mass (atleast of $10^{4}$ to $10^{5}$\(\textup{M}_\odot\) ) could provide seeds for the supermassive black holes in the galactic centre (GC), the generation of large scale-structure (LSS) through Poisson fluctuations \cite{lss} and effects in the thermal and ionisation history of the Universe. But the most compelling possibility is, these massive PBHs could account for all or a part of the dark matter (DM) density \cite{bertone,hooper,simpson,peebles,hui,khlopov,chapline}.\\*
 Due to their formation after inflation from large density fluctuations on small scales, PBHs represent a unique probe to study the small-scale early Universe, placing an upper limit on the primordial power spectrum spanning around 40 e-folds smaller than those visible in the Cosmic Microwave Background(CMB). The study of PBHs has also gained considerable importance in the light of the recent observations of gravitational waves (GWs) from merging black holes by Laser Interferometer Gravitational-Wave Observatory (LIGO)\cite{ligo1,ligo2,ligo3,ligo4}. The large masses of these black holes and the fact they have been observed at rather large distances suggest that they have formed relatively early in the Universe and therefore could be PBHs. Thus, studying the formation mechanism and the mass spectrum of PBHs more precisely could revolutionize our understanding of the Universe. \\*
 In this work, we will focus on the formation mechanism of PBHs via the dynamical evolution of a single inflaton field over a potential, inspired by string theory. Given the ultraviolet sensitivity of inflation and string theory being the best hope for
an ultraviolet theory including gravity, the study of the production mechanism of PBHs in string inspired inflationary models could allow us to probe the model parameters to the scales otherwise  cannot be studied through CMB.\\*
In this work, we refrain ourselves from detailed model building. Rather, with a phenomenologically motivated class of inflationary models with structural similarity to K\"ahler moduli inflation motivated models \cite{Sukanya1,mayukh,suru,conlon,Maharana:2015saa,bond,conlon1,pillado,dayan,linde1,teukolsky,mcdonough}, we try to understand all the theoretical and observational demands to produce PBHs, staying in the premise of the single field models \cite{Sukanya1,mayukh,conlon,bond},\cite{kohri,peiris,green3,niemeyer,niemeyer2,musco,  cicoli1,muia,burgess,kane,itzhaki,gb2}.
\\*
The class of the potential we have considered here can be represented by:\\
\begin{equation}
V = V_{0} ( \phi^{-\frac{n}{3}}+  \phi^{\frac{n}{3}})(1 + e^{-x \phi^{n m}})^{q}
\label{pot}
\end{equation} 
Here, $x$,$n$ and $m$ are the class defining set  parameters and $V_0$ fixes the scale of inflation and $q$ can be $\pm 1$. For, $q= +1$ with particular choices of set parameters, one  can get back the inflationary potential of the form of K\"ahler moduli inflation\cite{Maharana:2015saa}. We have also studied the $q= -1$ case. For both cases, we have studied the formation mechanism of PBHs. 
The rest of the paper is organised as follows. In  section \ref{inf}, we will briefly review the standard inflationary parameter estimation for this class of model. In section \ref{anpbh}, we have reported the analysis behind the PBH mass spectrum calculation and quoted our findings. Finally, the conclusions are drawn in section~\ref{conc}.
% and it can be constrained from the $\mathcal{P_{R}}(k)$ and  other parameters can be constrained from the cosmological observables like $r$, $n_s$. Here we are considering two fixed values for the $n= 1$ and $ 2$. 
 \section{Analysis for the inflationary observables}   
 \label{inf}
 To have successful inflation for at least  $60$ e-folds, one needs to have the inflaton field to roll on its potential slowly enough. The standard practice to design and carry out a successful inflationary epoch depends on calculating the slow-roll parameters $\epsilon$ and $\eta$  which are related to the accelerated expansion of the Universe through the Hubble parameter. When these parameters are expressed in terms of the potential (to the first-order approximation of the Hubble slow-roll), in our case, they turn out to be as follows \cite{Baumann,liddle} :
%%%%%%%%%%%%%%%%%%%%
\begin{equation}
\epsilon = \frac{1}{2} \left( \frac{V'}{V} \right)^2 = \frac{\left((1+ e^{x \phi^{m n}}) n (\phi^{2 n/3}-1)- 3~m~n~q x \phi^{m n} (1+ \phi^{2 n/3}) \right)^2}{18 (1+ e^{x\phi^{mn}})^2 ~\phi^2 (1+ \phi^{2 n/3})^2}
\label{eps}
\end{equation}
    
\begin{gather*}
\eta = \frac{V''}{V} = \frac{1}{9 \phi ^2 \left(\phi ^{\frac{2 n}{3}}+1\right) \left(e^{x \phi ^{m n}}+1\right)^2}  \Big(( n \big( 3+n -3 \phi^{2n/3} + n \phi^{2n/3}   + 9 mqx\phi^{mn}  \nonumber\\ +6 mnqx \phi^{mn} - 9 m^2 nqx \phi^{mn}+9 m^2 n q^2 x^2 \phi^{2mn} - 3m (-3+(2+3m)n)qx \phi^{(\frac{2}{3}+m)n} \nonumber\\ + 9 m^2nq^2 x^2 \phi^{\frac{2n}{3}+2mn} + e^{2x \phi^{mn}} +e^{2x\phi^{mn}}(3+n+(-3+n)\phi^{2n/3}) \nonumber\\
+e^{x \phi^{mn}} \big( 6-6 \phi^{2n/3}+9mqx \phi^{mn} (1+\phi^{2n/3})+ n \big( 2 + 2 \phi^{2n/3}-3m(-2+3m)qx\phi^{mn}+9m^2 qx^2 \phi^{2mn} \nonumber\\ - 3m(2+3m)qx\phi^{(\frac{2}{3}+m)n}+9m^2 q x^2 \phi^{\frac{2n}{3}+2mn}  \big) \big) \big) \Big)
\end{gather*}        

%%%%%%%%%%%%%%%%%%%%
Here prime denotes the derivative with respect to $\phi$, the inflaton field.
The amount of inflation is described in terms of the number of e-folds ($N_e$) during the inflationary epoch and in our case is given by:
%%%%%%%%%%%%%%%%%%%%%
\begin{equation}
N = {\int_{\phi _e}^{\phi _0} \frac{V}{{V\;'}} \, d\phi }={\int_{\phi _e}^{\phi _0} \frac{3 (1+ e^{x \phi ^{m n}}) \phi (1+ \phi^{2 n/3})}{(1+e^{x \phi^{m n}})~n (-1+\phi^{2n/3})-3 ~m~n~q~x \phi^{mn} (1+\phi^{2 n/3})} \, d\phi }
\label{efold}
\end{equation}

Using $\epsilon=1$ as a condition at the end of inflation, one can evaluate the $\phi_e$, where $\phi_0$ is the field value at the pivot ($k=0.05 Mpc^{-1}$). The scalar spectral index ($n_s$) and the tensor to scalar ratio ($r$) can be defined as:
\begin{equation}
n_s = 1 - 6 \epsilon + 2 \eta~~,~~~~~r = 16\epsilon~~.
\label{infobs}
\end{equation}
%%%%%%%%%%%%%%%%%%%%%%
The scalar power spectrum is defined as:
\begin{equation}
{P_\mathcal{R}}=\frac{1}{24 \pi^2}\left(\frac{V}{\epsilon}\right) = \frac{3(1+ e^{-x \phi^{mn}})^q V_{0} \phi^{2-n/3}(1+e^{x \phi^{mn}})^2(1+\phi^{2n/3})^3}{ 4 n^2 \pi^2\left( (1+e^{x \phi^{mn}})(\phi^{2n/3}-1)-3mxq\phi^{mn} (1+\phi^{2n/3})  \right)^2}
\label{amplitude}
\end{equation}

\subsection{Inflationary Observables for $q=-1$}

Considering $q=-1$ and keeping $n=1$ in Eq.(\ref{pot}) and choosing the suitable value of other  potential parameters ($x$ and $m$),  we can calculate the tensor to scalar ratio($r$) and the spectral index($n_s$). However, due to the sophisticated nature of the potential, it is not possible to solve the Eq. (\ref{efold}) analytically, here we use the numerical approach for calculating the inflationary observables. Using Eq. (\ref{eps}),(\ref{efold}) and (\ref{infobs}), we calculate the inflationary observables:

\begin{center}
\begin{table*}[h]
\begin{center}
% \begin{tabular}{|l|l |r| l|  }
% \hline
% \multicolumn{1}{|c|}{ {$x$}} & \multicolumn{1}{c|}{ $m$ } & \multicolumn{1}{c|}{ $r$ }& \multicolumn{1}{c|}{$n_s$}\\
% \hline
% \,\,\,\,\,\,$0.45018$\,\,\,\,& \,\,$3/5$\,\, &\,\,  $0.0490$\,\,& \,\,$0.9680$
% \\
% \,\,\,\,\,\,$0.6140$\,\,\,\,& \,\,$1/2$\,\, &\,\, $0.0451$\,\,& \,\,$0.9682$
% \\
% ~~~$0.50117$\,\,\,\,& \,\,$2/5$\,\, &\,\, $ 0.0535$\,\,& \,\,$0.9719$

%  \\
 
% \hline
% \end{tabular}
\begin{tabular}{|c|c|c|c|c|c|c|}
\hline
   & \multicolumn{2}{c|}{$x=0.45018, m=3/5$} & \multicolumn{2}{c|}{$x=0.6140, m=1/2$} & \multicolumn{2}{c|}{$x=0.50117, m=2/5$} \\ \hline
$N$  & $n_s$               & $r$                & $n_s$              & $r$                & $n_s$              & $r$                 \\ \hline
55 & 0.9654            & 0.0559           & 0.9642            & 0.0518           & 0.9818           & 0.0584           \\ \hline
60 & 0.9680             & 0.0490           & 0.9680            & 0.0451           & 0.9710            & 0.0530            \\ \hline
65 & 0.9710             & 0.0435           & 0.9710            & 0.0390           & 0.9700            & 0.0477            \\ \hline
70 & 0.9743             & 0.0390           & 0.9740            & 0.035            & 0.9722            & 0.0426            \\ \hline
\end{tabular}
\end{center}
\caption{For $q=-1, n=1$, Inflationary observables for the different choices of the potential parameters ($x$ and $m$) with variety of total number of e-fold, all the observables are consistent with the $Planck'18$ \cite{planck'18} }
\label{tab1}
\end{table*}
\end{center}

\begin{figure}[htb!]
%\label{N_e_nsr}
  \centering
  \subfloat[]{\includegraphics[height= 6cm, width=8.1cm]{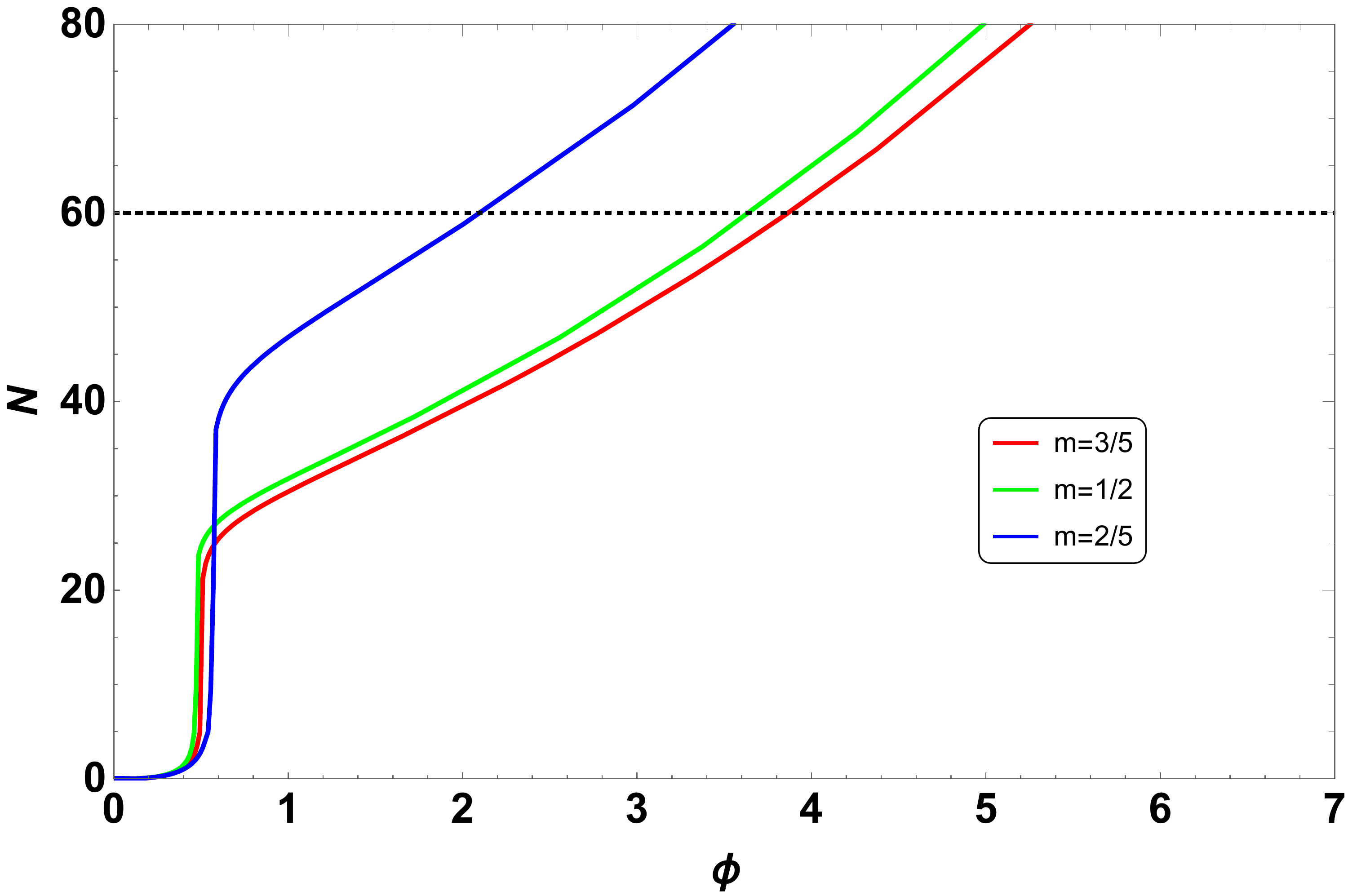}\label{NvsPhi-1}}
  \hfill
  \subfloat[]{\includegraphics[height= 6cm, width=8.1cm]{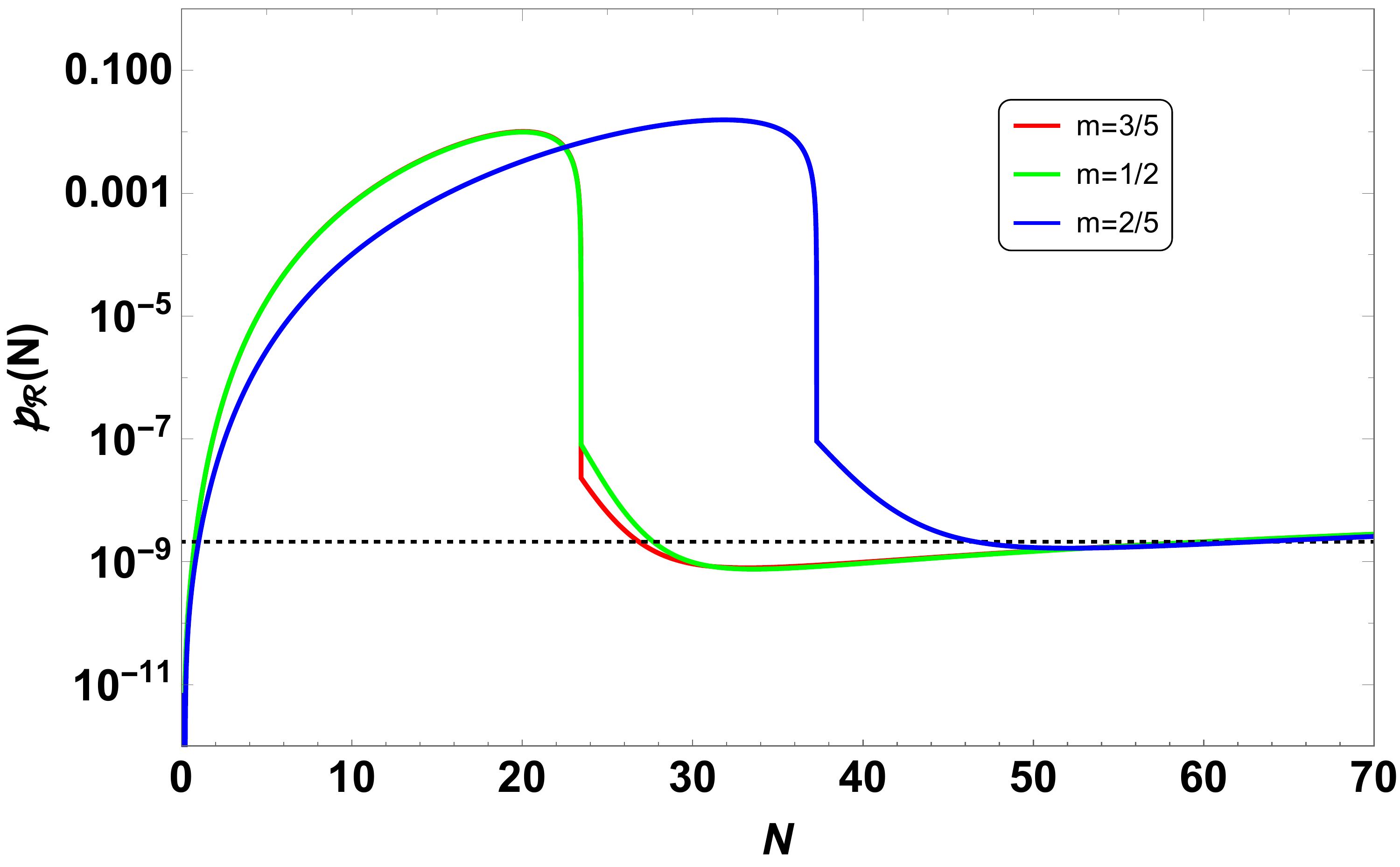}\label{PrvsN-1}}
  \hfill
 
\caption{\small{ {\bf Left Panel:} Three different colors correspond to three different values of potential parameter $x$. Red color corresponds to $x=0.45018$, green is for $x=0.6140$ and blue stands for $x=0.50117$. The Number of e-folds ($N$) as a function of the scalar field ($\phi$). The dotted black line corresponds to the $N=60$.\\
{\bf Right Panel:} The scalar power spectrum as a function of the Number of e-folds ($N$). Three different colors correspond to three different values of potential parameter $x$. Red color corresponds to $x=0.45018$, green is for $x=0.6140$ and blue stands for $x=0.50117$. Peak in the ${P_\mathcal{R}}$  can be seen clearly which is necessary for the production of the PBHs.  The black dotted line shows the value of ${P_\mathcal{R}}$ ($= 2.1\times10^{-9}$) at large scales and it is in good agreement with the $Planck'18$\cite{planck'18}. Enhancement in the power spectrum is calculated numerically using Eq. (\ref{amplitude})}}
\label{plot1}
\end{figure}
 
Keeping $n=2$  and taking different values of potential parameters $x$ and $m$ and  following the same approach, using Eq. (\ref{eps}),(\ref{efold}) and (\ref{infobs}), one can calculate the inflationary observables:

\begin{center}
\begin{table*}[ht]
\begin{center}
\begin{tabular}{|c|c|c|c|c|c|c|}
\hline
   & \multicolumn{2}{c|}{$x=6.0, m=-1/2$} & \multicolumn{2}{c|}{$x=8.4, m=-1/2$} & \multicolumn{2}{c|}{$x=8.6, m=-3/5$} \\ \hline
$N$  & $n_s$             & $r$               & $n_s$             & $r$               & $n_s$             & $r$               \\ \hline
55 & 0.9692           & 0.0323          & 0.9403           & 0.0548          & 0.9449           & 0.0251          \\ \hline
60 & 0.9756           & 0.0287          & 0.9519           & 0.0432          & 0.9596           & 0.0201          \\ \hline
65 & 0.9808           & 0.0262          & 0.9612           & 0.0356          & 0.9704           & 0.0171          \\ \hline
70 & 0.9847           & 0.0245          & 0.9682           & 0.0305          & 0.9783           & 0.0152          \\ \hline
\end{tabular}
\end{center}
\caption{For $q=-1, n=2$, Inflationary observables for the different choices of the potential parameters ($x$ and $m$) with variety of total number of e-folds, all the observables are consistent with the $Planck'18$ \cite{planck'18} }
\label{tab1}
\end{table*}
\end{center}
%\subsection{Inflationary Observables for $n=2$}

% \begin{center}
% \begin{table*}[ht]
% \begin{center}
% \begin{tabular}{|l|l |r| l|  }
% \hline
% \multicolumn{1}{|c|}{ {$x$}} & \multicolumn{1}{c|}{ $m$ } & \multicolumn{1}{c|}{ $r$ }& \multicolumn{1}{c|}{$n_s$}\\
% \hline
% \,\,\,\,\,\,$6.0$\,\,\,\,& \,\,$-1/2$\,\, &\,\,  $0.0287$\,\,& \,\,$0.9755$
% \\
% \,\,\,\,\,\,$8.4$\,\,\,\,& \,\,$-1/2$\,\, &\,\, $0.0370$\,\,& \,\,$0.9594$
% \\
% ~~~$8.6$\,\,\,\,& \,\,$-3/5$\,\, &\,\, $ 0.0201$\,\,& \,\,$0.9596$

%  \\
 
% \hline
% \end{tabular}

% \end{center}
% \caption{Inflationary observables for the different values of the potential parameters ($x$ and $m$).All the observables are consistent with the $Planck'18$ \cite{planck'18}}
% \label{tab1}
% \end{table*}
% \end{center}

\begin{figure}[htb!]
%\label{N_e_nsr}
  \centering
  \subfloat[]{\includegraphics[height= 6cm, width=8.21cm]{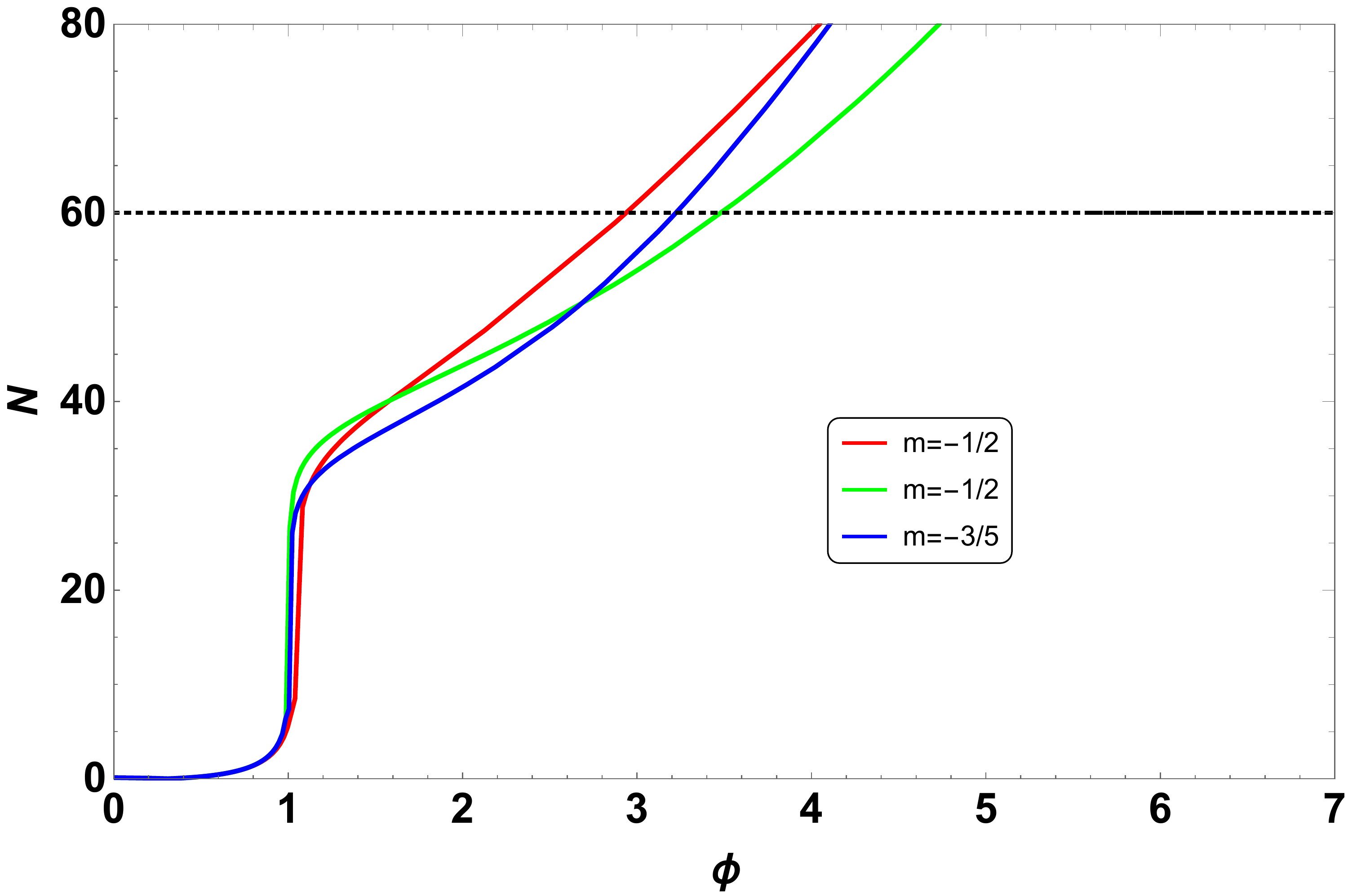}\label{NvsPhi-1}}
  \hfill
  \subfloat[]{\includegraphics[height= 6cm, width=8.21cm]{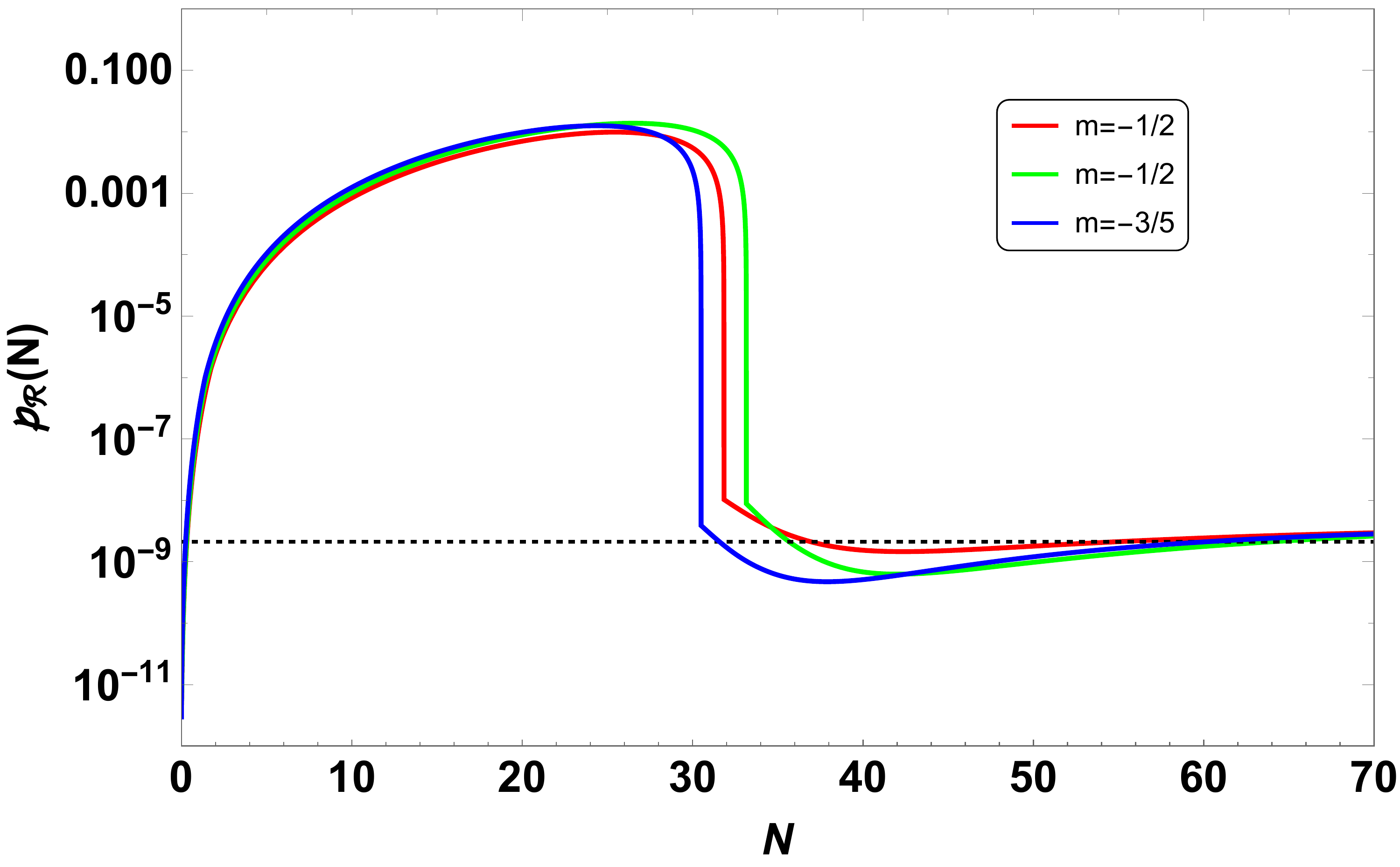}\label{PrvsN-1}}
  \hfill
 
\caption{ \small{ {\bf Left Panel:} Three different colors correspond to three different values of potential parameter $x$. Red color corresponds to $x=8.4$, green is for $x=6.0$ and blue stands for $x=8.6$. The number of e-folds ($N$) as a function of the scalar field ($\phi$). The dotted black line corresponds to the $N=60$.\\
{\bf Right Panel:} The scalar power spectrum as a function of the Number of e-folds ($N$). Three different colors correspond to three different values of potential parameter $x$. Red color corresponds to $x=8.4$, green is for $x=6.0$ and blue stands for $x=8.6$. Peak in the ${P_\mathcal{R}}$  can be seen clearly which is necessary for the production the PBHs.  The black dotted line shows the value of ${P_\mathcal{R}}$ ($= 2.1\times10^{-9}$) at large scales and it is in good agreement with the $Planck'18$ \cite{planck'18}. Enhancement in the power spectrum is calculated numerically using Eq. (\ref{amplitude})}}
\label{plot2}
\end{figure}

\subsection{Inflationary Observables for $q=1$}
Using the same approach as we did in the last section, setting $q=1$ and $n=1$ in Eq. (\ref{pot}-\ref{infobs}) and for different values of potential parameters $x$ and $m$, we can compute the tensor to scalar ratio and the spectral index 
\begin{center}
\begin{table*}[ht]
\begin{center}
% \begin{tabular}{|l|l |r| l|  }
% \hline
% \multicolumn{1}{|c|}{ {$x$}} & \multicolumn{1}{c|}{ $m$ } & \multicolumn{1}{c|}{ $r$ }& \multicolumn{1}{c|}{$n_s$}\\
% \hline
% \,\,\,\,\,\,$-2.14990$\,\,\,\,& \,\,$1/25$\,\, &\,\,  $0.04540$\,\,& \,\,$0.97035$
% \\
% \,\,\,\,\,\,$-1.69995$\,\,\,\,& \,\,$1/20$\,\, &\,\, $0.04449$\,\,& \,\,$0.97338$
% \\
% ~~~$-1.17090$\,\,\,\,& \,\,$3/50$\,\, &\,\, $ 0.03336$\,\,& \,\,$0.98122$

%  \\
 
% \hline
% \end{tabular}
\begin{tabular}{|c|c|c|c|c|c|c|}
\hline
   & \multicolumn{2}{c|}{$x=-2.1499, m=1/25$} & \multicolumn{2}{c|}{$x=-1.6999, m=1/20$} & \multicolumn{2}{c|}{$x=-1.1709, m=3/50$} \\ \hline
$N$  & $n_s$               & $r$                  & $n_s$                & $r$                 & $n_s$               & $r$                  \\ \hline
55 & 0.9731             & 0.0509             & 0.9800             & 0.0487           & 0.9864            & 0.0354            \\ \hline
60 & 0.9703             & 0.0454             & 0.9734             & 0.0445           & 0.9812            & 0.0333             \\ \hline
65 & 0.9715            & 0.0403             & 0.9733             & 0.0399           & 0.9798            & 0.0308             \\ \hline
70 & 0.9735             & 0.0359            & 0.9743            & 0.03580           & 0.9799             & 0.0284            \\ \hline
\end{tabular}
\end{center}
\caption{For $q=1$ and $n=1$, Inflationary observables for the different values of the potential parameters ($x$ and $m$) with variety of total number of e-folds.}
\label{tab1}
\end{table*}
\end{center}

\begin{figure}[htb!]
%\label{N_e_nsr}
  \centering
  \subfloat[]{\includegraphics[height= 6cm, width=8.21cm]{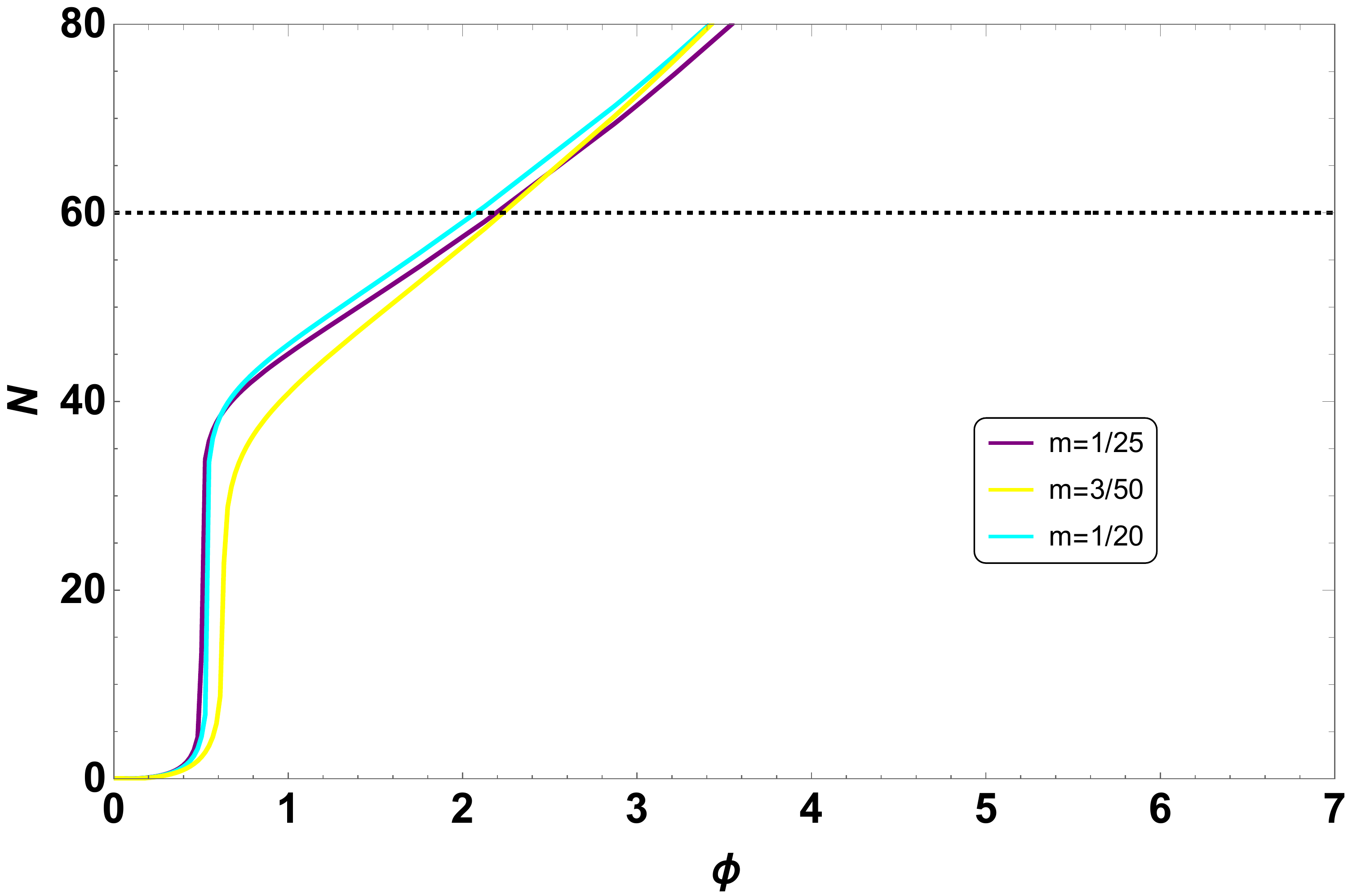}\label{NvsPhi-1}}
  \hfill
  \subfloat[]{\includegraphics[height= 6cm, width=8.21cm]{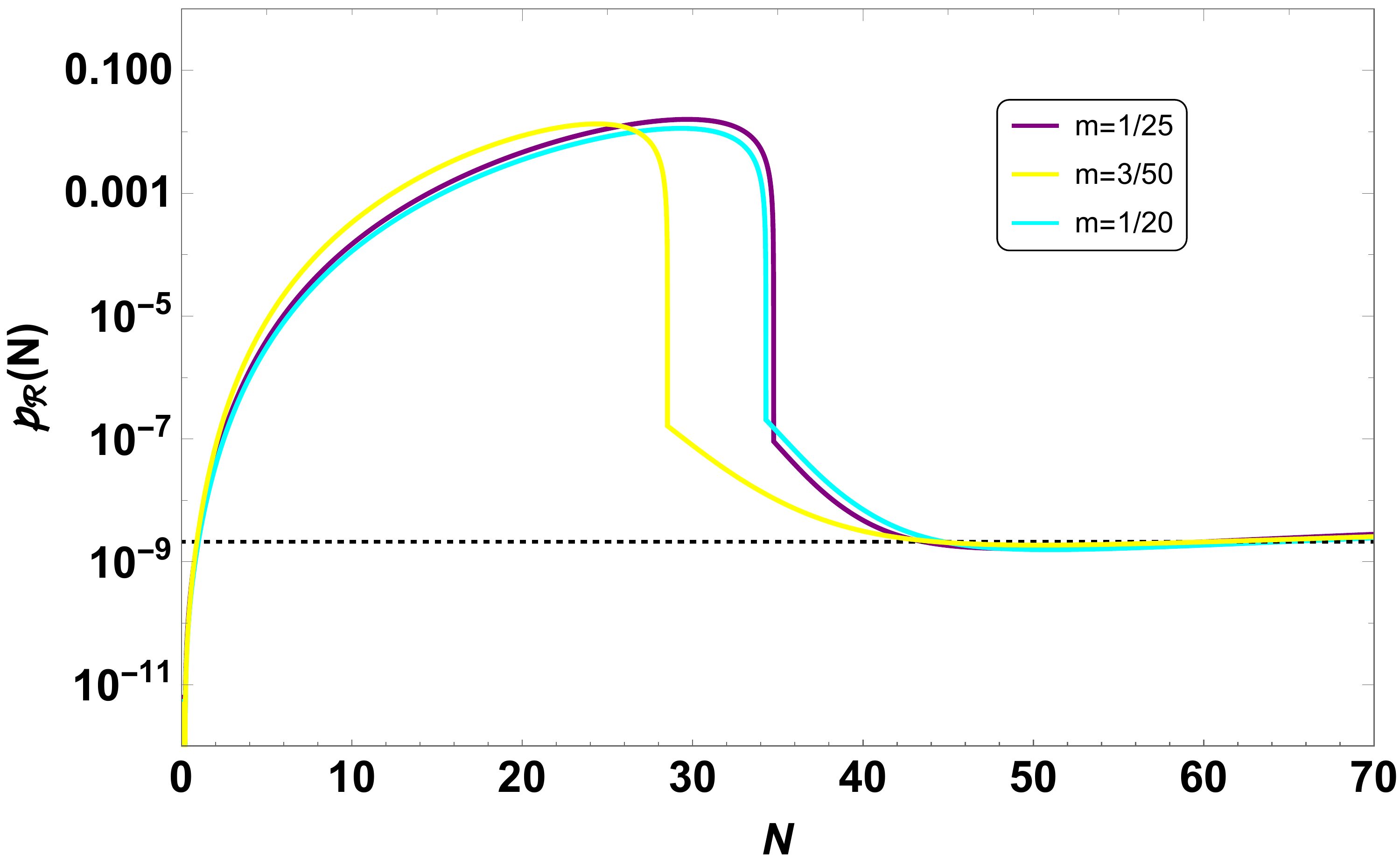}\label{PrvsN-1}}
  \hfill
 
\caption{\small{ {\bf Left Panel:}  Three different colors correspond to three different values of potential parameter $x$. Purple color corresponds to $x= -2.1499$, yellow is for $x= -1.1709$ and cyan stands for $x=-1.6999$. The number of e-folds ($N$) as a function of the scalar field ($\phi$). The dotted black line corresponds to the $N=60$.\\
{\bf Right Panel:} The scalar power spectrum as a function of the Number of e-folds ($N$).  Three different colors correspond to three different values of potential parameter $x$. Purple color corresponds to $x= -2.1499$, yellow is for $x= -1.1709$ and cyan stands for $x=-1.6999$. Peak in the ${P_\mathcal{R}}$  can be seen clearly which is necessary for the production the PBHs.  The black dotted line shows the value of ${P_\mathcal{R}}$ ($= 2.1\times10^{-9}$) at large scales and it is in good agreement with the $Planck'18$ \cite{planck'18}, enhancement in the power spectrum is calculated numerically using Eq. (\ref{amplitude})}}
\label{plot2}
\end{figure}

Setting $q=1$ and $n=2$ in Eq. (\ref{pot}-\ref{infobs}) and for different values of potential parameters $x$ and $m$, we can compute the tensor to scalar ratio and the spectral index : 

\begin{center}
\begin{table*}[ht]
\begin{center}
\begin{tabular}{|c|c|c|c|c|c|c|}
\hline
   & \multicolumn{2}{c|}{$x=-1.5432, m=1/25$} & \multicolumn{2}{c|}{$x=-1.1222, m=3/100$} & \multicolumn{2}{c|}{$x=-1.4500, m=1/100$} \\ \hline
$N$  & $n_s$               & $r$                 & $n_s$                & $r$                 & $n_s$               & $r$                  \\ \hline
55 & 0.9593            & 0.0983            & 0.9542              & 0.1067            & 0.9564             & 0.0993            \\ \hline
60 & 0.9642            & 0.0860            & 0.9600             & 0.0917            & 0.9616             & 0.0857            \\ \hline
65 & 0.9680            & 0.0764            & 0.9650             & 0.0790            & 0.9666             & 0.0734             \\ \hline
70 & 0.9711            & 0.0687            & 0.9684              & 0.0711            & 0.9694             & 0.0668             \\ \hline
\end{tabular}
\end{center}
\caption{For $q=1$ and $n=2$, Inflationary observables for the different choices of the potential parameters ($x$ and $m$) with variety of total number of e-folds.}
\label{tab1}
\end{table*}
\end{center}

\begin{figure}[htb!]
%\label{N_e_nsr}
  \centering
  \subfloat[]{\includegraphics[height= 6cm, width=8.21cm]{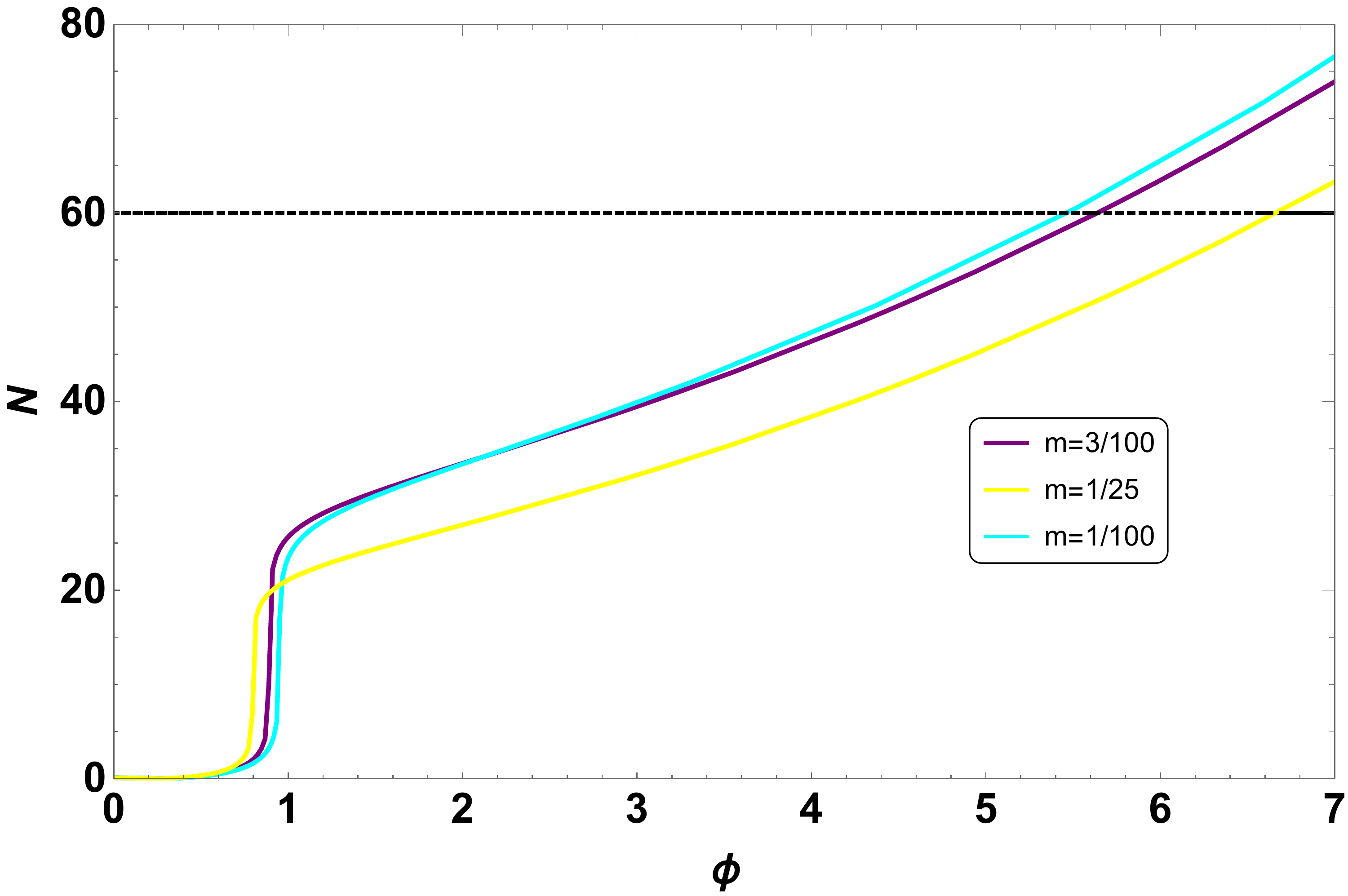}\label{NvsPhi-1}}
  \hfill
  \subfloat[]{\includegraphics[height= 6cm, width=8.21cm]{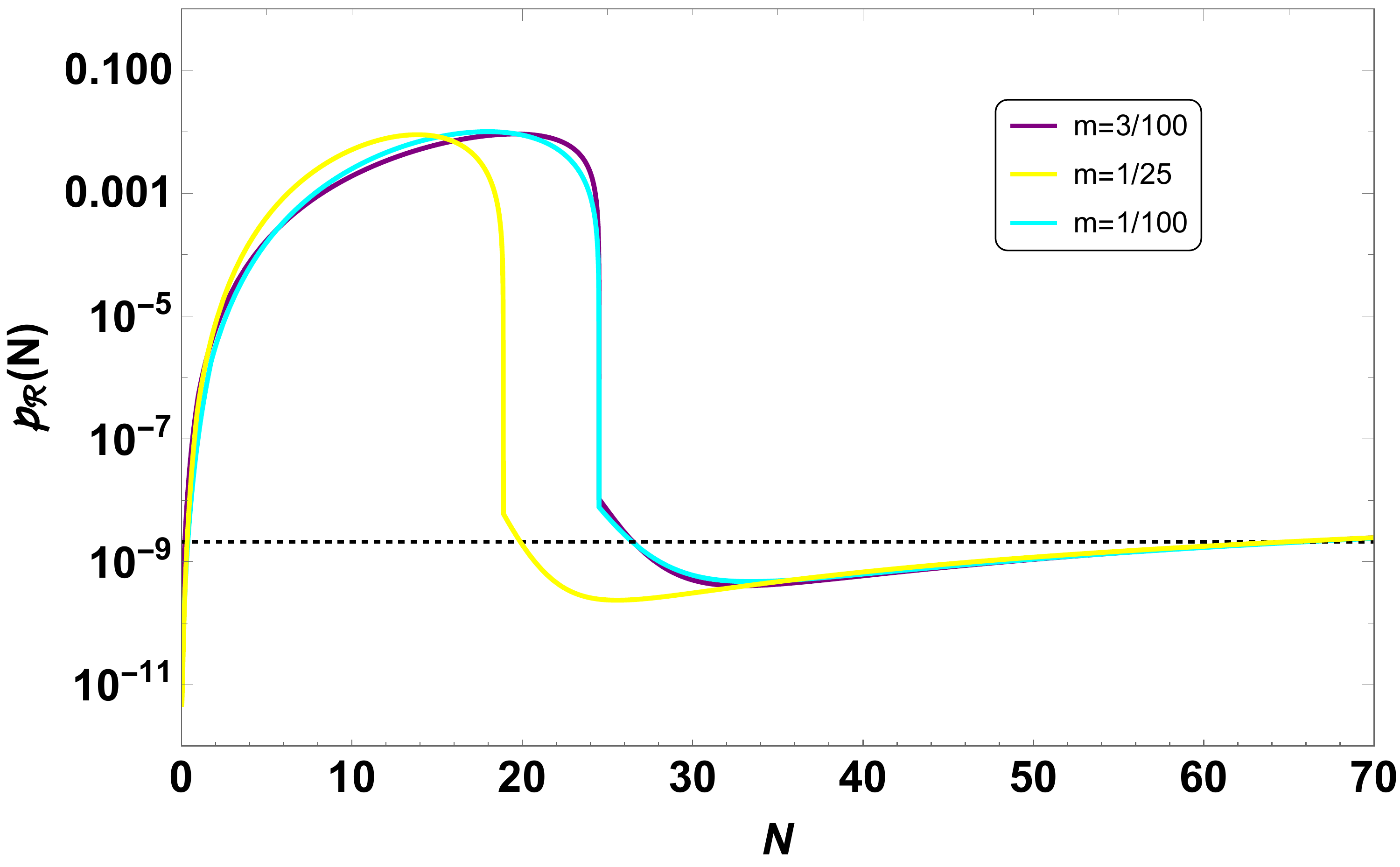}\label{PrvsN-1}}
  \hfill
 
\caption{\small{ {\bf Left Panel:} Three different colors correspond to three different values of potential parameter $x$. Purple color corresponds to $x= -1.1222$, cyan is for $x= -1.45$ and yellow stands for $x=-1.5432$. The Number of e-folds ($N$) as a function of the scalar field ($\phi$). The dotted black line corresponds to the $N=60$. \\
{\bf Right Panel:} The scalar power spectrum as a function of Number of e-folds ($N$). Three different colors correspond to three different values of potential parameter $x$. Purple color corresponds to $x= -1.1222$, cyan is for $x= -1.45$ and yellow stands for $x=-1.5432$. Peak in the ${P_\mathcal{R}}$  can be seen clearly which is necessary for the production the PBHs.  The black dotted line shows the value of ${P_\mathcal{R}}$ ($= 2.1\times10^{-9}$) at large scales and it is in good agreement with the $Planck'18$ \cite{planck'18}. Enhancement in the power spectrum is calculated numerically using Eq. (\ref{amplitude})}}
\label{plot2}
\end{figure}

\section{Analysis of PBH Production}
\label{anpbh}
\noindent
%%%%%%%%%%%%%
\begin{figure}[htb!]
%\label{N_e_nsr}
\centering
\subfloat[]{\includegraphics[height= 6cm, width=8.21cm]{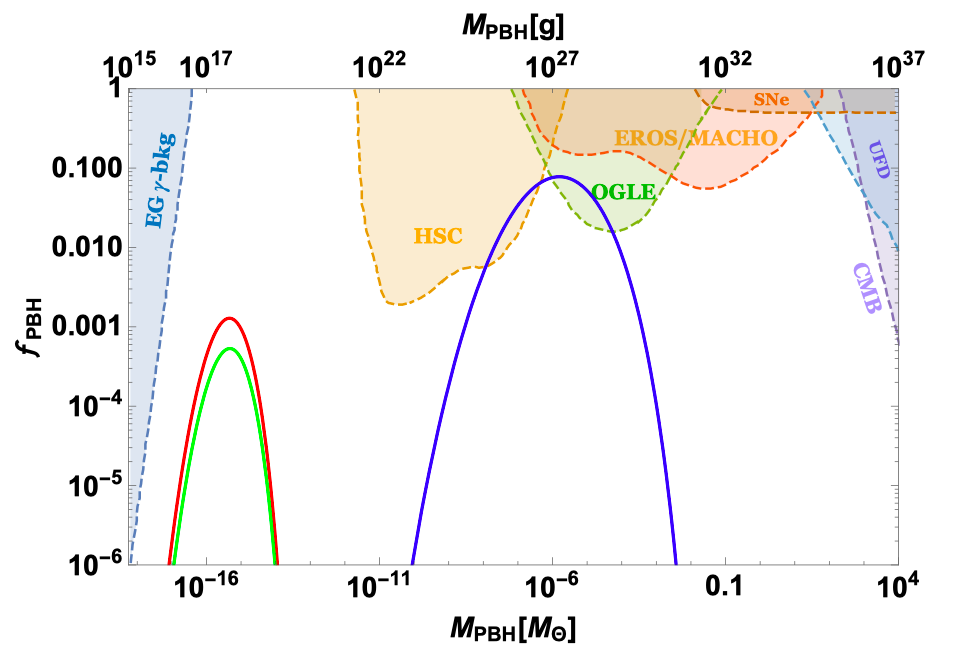}\label{fpbhvsmpbh-1}}
  \hfill
\subfloat[]{\includegraphics[height= 6cm, width=8.21cm]{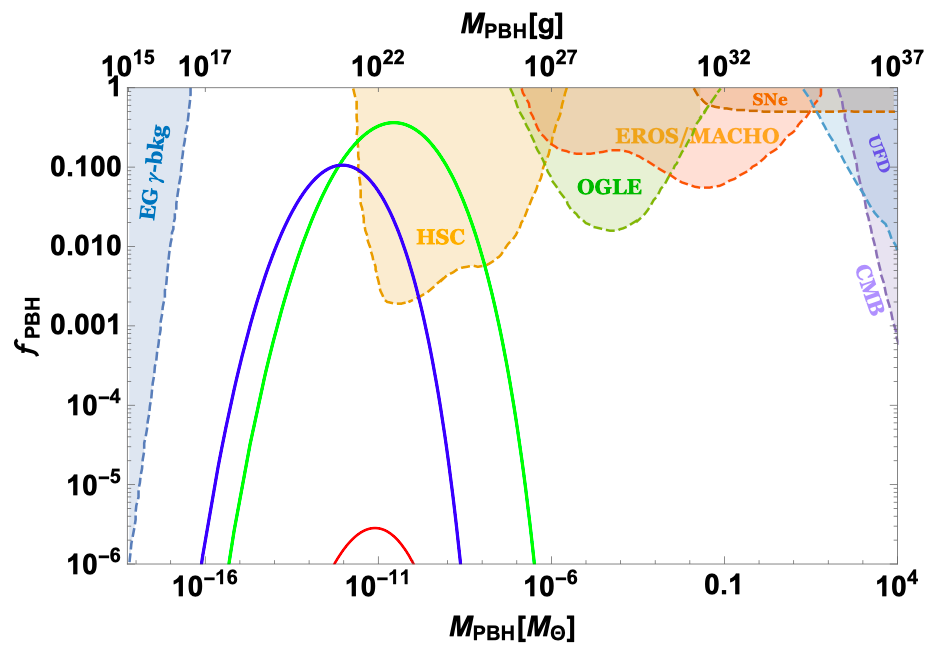}\label{fpbhvsmpbh-2}}
  \hfill
 
\caption{\small{ {\bf Left Panel:} For $q=-1$ and $n=1$, Dark matter fraction ($f_{PBH}$) as a function of the primordial black hole mass ($M_{PBH}$) in the unit of \(\textup{M}_\odot\) is plotted. Three different colors correspond to three different values of potential parameter $x$ and $m$. Red color corresponds to $x=0.45018,~m =3/5$, green is for $x=0.6140~,m=1/2$ and blue stands for $x=0.50117,~m= 2/5$.\\
 {\bf Right Panel:} For  $q=-1$ and  $n=2$, Dark matter fraction ($f_{PBH}$) as a function of the primordial black hole mass ($M_{PBH}$) in the unit of \(\textup{M}_\odot\) is plotted. Three different colors correspond to three different values of potential parameter $x$ and $m$. Color red corresponds to $x=6.0,~m =-1/2$, green is for $x=8.4~,m=-1/2$ and blue stands for $x=8.6,~m= -3/5$. }}
\label{plot3}
\end{figure}
%%%%%%%%%%%
The large primordial scalar fluctuations which are generated during the inflationary epoch can give rise to the large density fluctuations which plays the crucial role in the formation of the primordial black holes.

% A peak in the scalar power spectrum at large scale can give rise to the production of primordial black holes when the large density perturbation collapse in the early universe it can produce significant amount of Primordial Black Holes. 
\noindent
PBHs have been the interest of discussion for many years. A significant or whole amount of Dark Matter quantity can be attributed to the PBHs. The abundance of the primordial black holes is very tightly constrained due to their gravitational effects and from their rate of evaporation. The amplitude of the curvature power spectrum  required for the production of  PBHs in the early universe is of order $10^{-3}-10^{-2}$\cite{gb2}. The inflationary curvature perturbation ($\zeta$) and the density contrast ($\Delta \rho= \frac{\delta \rho}{\rho}$) are related as \cite{Kalaja,Sukanya,musco1,musco2}: 
  
  \begin{equation}
	\delta ({\bf x},t)=\frac{2(1+w)}{5+3w}\bigg( \frac{1}{aH}\bigg)^2\bigtriangledown ^2\zeta ({\bf x},t),
	\label{delta1}
\end{equation}
\noindent
Here $a$ is the scale factor and $w$ is the equation of state. When the density fluctuation is higher than the critical density($\delta_c$), the overdense region collapses and forms the PBHs. The critical density parameter is related to the equation of state of the background as \cite{Jedamzik,Musko,Harada}:
\begin{equation}
	\delta_c = \dfrac{3(1+w)}{5+3w}\sin^2\left(\dfrac{\pi\sqrt{w}}{1+3w}\right)\,.\label{deltac1}
\end{equation}
For the radiation dominated epoch ($w=1/3$) the critical density parameter can be calculated which is $\delta_c= 0.414$. One may note that Eq. (\ref{deltac1}) is not valid form the matter dominated epoch where $\omega=0$. The mass fraction ($\beta(M)$) of a PBH of mass $M$ can be formulated using the Press-Schechter formalism

\begin{equation}\label{eq:beta}
	\beta(M) = \int_{\delta_c}^{\infty} d\delta ~P(\delta) = \frac{2}{\sqrt{2\pi}\sigma(M)} \int^\infty_{\delta_c} d\delta \exp\left(-\frac{\delta^2}{\sigma(M)^2}\right) = \mathrm{erfc}\left(\frac{\delta_c}{\sqrt{2}\sigma(M)}\right)\,.
\end{equation} 
\noindent
The primordial curvature perturbations ($P_{\zeta}(k)$)  and the variance of the density fluctuations ($\sigma^{2}(M)$) are related as: 
\begin{equation}
	\sigma^2(M)=\frac{4(1+w)^2}{(5+3w)^2}\int \frac{dk}{k}(kR)^4 W^2(k,R)P_{\zeta}(k),\label{sigmaM1}
\end{equation}
Here the window function $W(k,R)$  can be chosen to be the Gaussian function. Using the  following approximation we can write a more simplified version of the variance of the density fluctuations \cite{Sukanya} 
\begin{equation}
	\sigma (M) \simeq \frac{2(1+w)}{(5+3w)}\sqrt{P_{\zeta}(k)}
	\label{sigmaM2}
\end{equation}
\noindent
The mass of PBH at the formation can be written as a fraction of the horizon mass
\begin{equation}
M_{PBH} = \gamma \frac{4 \pi  M^2_p}{H_N} e^{2N}
\label{Mpbh}
\end{equation}
 Here $\gamma$ is the efficiency factor, we take $\gamma\sim0.4$ \cite{gb1,gb2} and $N$ is the number of e-folds during horizon exit and $H_{N}$ is the Hubble expansion rate near the inflection point. The value of $H_N$ depends upon the potential parameters($x,m$) and class defining parameters ($q,n$). The PBH mass fraction at the formation can be related to the present PBH density parameter($\Omega^0_{PBH}$) as: 

\begin{equation}
\beta(M_{\rm PBH}) \equiv \frac{\rho_{\rm PBH}^{\rm i}}{\rho_{\rm crit}^{\rm i}}
=\frac{\rho_{\rm PBH}^{\rm eq}}{\rho_{\rm crit}^{\rm eq}}\left(\frac{a_{\rm i}}{a_{\rm eq}}\right)
\approx \Omega_{\rm PBH}^{0} \left(\frac{a_{\rm i}}{a_{\rm eq}}\right)\,,
\end{equation}
Here, $a_{eq}$ refers to the scale factor at the matter-radiation equality, $a$ is the scale factor and $\rho_{crit}$ is the critical energy density. We have the relation $s= g_{\ast,s} a^3 T^3$, where $g_\ast$ is the number of degrees of freedom at constant entropy. From radiation density $\rho= \frac{\pi^2}{30} g_\ast  T^4$ and the horizon mass $M_H= \frac{4 \pi}{3}\rho H^{-3}$, we obtain: 

\begin{equation}
\beta(M_{\rm PBH}) = \Omega_{\rm PBH}^{0} \left( \frac{g_{\star}^{\rm eq}}{g_{\star}^{\rm i}} 
   \right)^{1/12}  \left( \frac{M_{\rm H}}{M_{\rm H}^{\rm eq}} \right)^{1/2} \,,
\end{equation}

The horizon mass at matter radiation equality can be written as \cite{Green, Karim}
\begin{equation}
M_{\rm H}^{\rm eq} = \frac{4 \pi}{3} \rho_{\rm eq} H_{\rm eq}^{-3}
          = \frac{8 \pi}{3} \frac{\rho_{\rm rad}^{0}}{a_{\rm eq} k_{\rm eq}^3} \,.
\end{equation} 
Considering $g_\ast=g_{\ast,s}$ and inserting all the numerical values \cite{WMAP,Karim},   $g_{\star}^{\rm eq}
\approx 3$, $g_{\star}^{\rm i} \approx 100$,$\Omega_{\rm rad}^{0} h^2 = 4.17 \times 10^{-5}$,
$\rho_{\rm crit}  = 1.88 \times 10^{-29}  h^2 \, {\rm g \, cm}^{-3}$,
$k_{\rm eq} = 0.07 \, \Omega_{\rm m}^{0} h^2 \, {\rm Mpc}^{-1}$, $a_{\rm
eq}^{-1} = 24 000 \, \Omega_{\rm m}^{0} h^2$ and $\Omega_{\rm m}^{0} h^2 =
0.1326 \pm 0.0063$ gives $M_{\rm H}^{\rm eq}= 1.3 \times
10^{49} (\Omega_{\rm m} h^2)^{-2} \, {\rm g}$, we obtain
\begin{equation}
\label{betaomega}
\beta(M_{\rm {PBH}}) = 6.4 \times 10^{-19} \, \Omega_{\rm PBH}^{0} 
   \left( \frac{M_{\rm PBH}}{\gamma \times  5\times10^{14} \, {\rm g}} \right)^{1/2} \,.
\end{equation}
The PBHs having mass less than $5\times10^{14}g$ cannot contribute to the present dark matter fraction, as they would have been evaporated by today. The mass of PBHs produced depends on the value of $H_N$ at the inflection point. From Eq. (\ref{Mpbh}), we can establish a relation between the mass of produced black holes and the number of e-folds. For  $q=-1$ and $n=1$ and $x=0.45018$, $m=3/5$ and using Eq. (\ref{Mpbh}), we obtain \cite{gb1,gb2}

\begin{equation}
M_{PBH}= e^{2(N-37.0119)}
\label{mass1}
\end{equation}
\noindent
Using  the Eq. (\ref{Mpbh}) and Eq. (\ref{mass1}) it is straightforward to calculate the $M_{PBH}$ for the different combinations of class defining parameters ($n,q$) and for the potential parameters ($x,m$)  Since it is not possible to solve the Eq.(\ref{efold}) analytically we use numerical approach and from Eq.(\ref{amplitude}) we can establish a relation between power spectrum and number of e-folds using the same numerical methods. This has been shown in the figure (\ref{PrvsN-1}). From Eq. (\ref{mass1}) we can relate $M_{PBH}$ and $P_{\zeta}(k)$. We know that the variance$(\sigma(M))$ of density fluctuation depends on the power spectrum hence, using Eq. (\ref{eq:beta}),(\ref{sigmaM2})and  Eq.(\ref{betaomega}) we can write the relation between present dark matter fraction $(f_{PBH}= \frac{\Omega^0_{PBH}}{\Omega^0_{DM}} )$ and $(M_{PBH})$. Dark matter fraction $(f_{PBH})$ depends on the amplitude of $P_{\zeta}(k)$ which means that even small fluctuations in the power spectrum can lead to abruptly small or large value of dark matter  fraction. Therefore, one needs to calculate the the $P_{\zeta}(k)$ precisely to get the $f_{PBH}$ in the correct order. Now, it is straightforward to calculate $f_{PBH}$ and $(M_{PBH})$ in the similar fashion for the different set of potential parameters $x,m$ and and class defining parameter $n$,  which is shown in the figure (\ref{plot3}). For the monochromatic mass distribution one can see the present bounds on PBHs \cite{constraints} coming from the Extragalactic Gamma Background $(EG~\gamma-bkg)$ \cite{gammaray}, $HSC$ \cite{hsc}, OGLE \cite{ogle}, EROS/MACHO \cite{eros,macho}, $SNe,UFD$ and $CMB$ \cite{carr5}.

%%%%%%%%%%%%%5

\begin{figure}[htb]
%\label{N_e_nsr}
\centering
\subfloat[]{\includegraphics[height= 6cm, width=8.21cm]{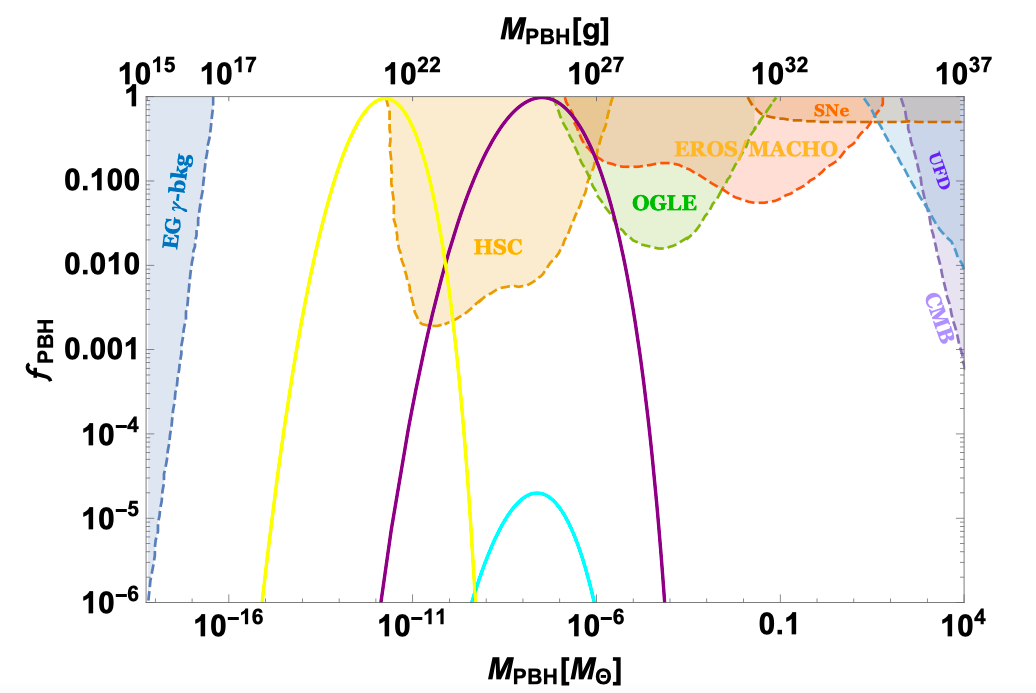}\label{fpbhvsmpbh-3}}
  \hfill
\subfloat[]{\includegraphics[height= 6cm, width=8.21cm]{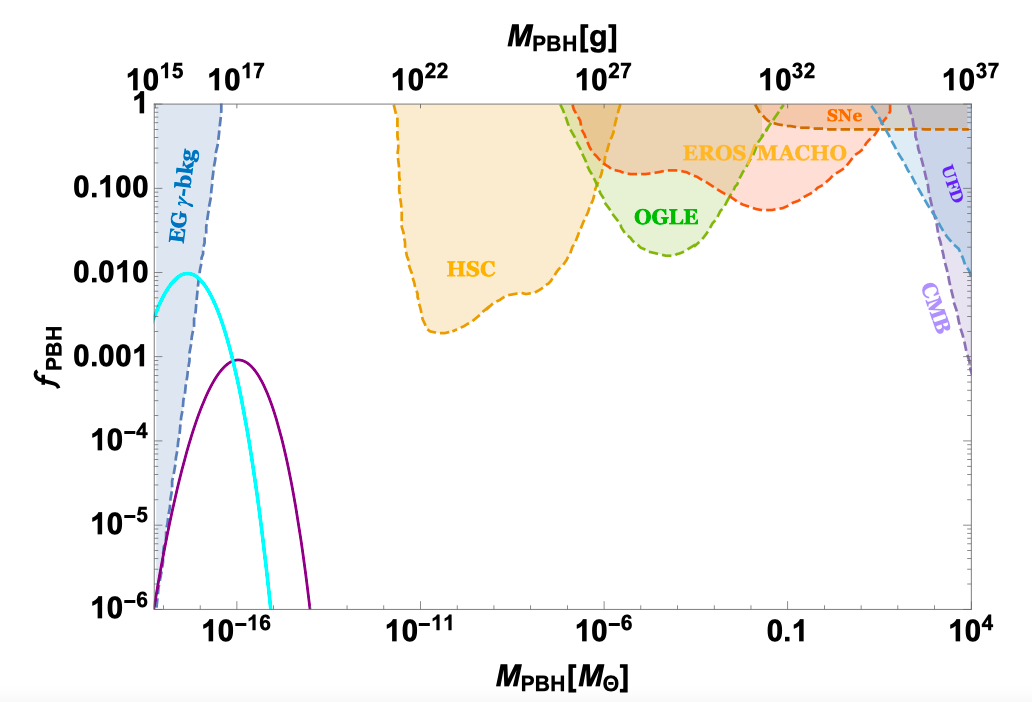}\label{fpbhvsmpbh-4}}
  \hfill
 
\caption{\small{ {\bf Left Panel:} For $q=1$ and $n=1$, Dark matter fraction ($f_{PBH}$) as a function of the primordial black hole mass ($M_{PBH}$) in the unit of  \(\textup{M}_\odot\) is plotted. Three different colors correspond to three different values of potential parameter $x,m$. Purple color corresponds to $x= -2.14990, m=1/25$, yellow is for $x= -1.17090, m=3/50$ and cyan stands for $x=-1.69995, m=1/20$.\\
 {\bf Right Panel:} For  $q=1$ and  $n=2$, Dark matter fraction ($f_{PBH}$) as a function of the primordial black hole mass ($M_{PBH}$) in the unit of  \(\textup{M}_\odot\) is plotted. Two different colors correspond to two different values of potential parameter $x$ and $m$. Purple color corresponds to $x= -1.1222, m=3/100$, cyan is for $x= -1.45, m=1/100$. }}
\label{plot4}
\end{figure}
%%%%%%%%%%%%%%%%

\section{Conclusion}
\label{conc}
In this work we have worked with a class of potential which has the correct feature of slow roll to ultra slow roll transition required to have an enhancement in the power spectrum which seeds the production of PBHs when the scales re-enters the horizon during later epochs. The model considered here is compatible with the constraints of $Planck'18$ on inflationary observables. For both $q= \pm 1$, we have shown, for certain choice of parameters, PBHs can be produced in different mass ranges from $10^{-18}$ to $10^{-4}$  \(\textup{M}_\odot\). The PBHs produced in both the subclass of this model can be attributed to the total dark matter density today from some percentage to $100 \%$. 

\underline{$q= -1$ case:}

 We have done the inflationary and the PBHs production analysis for the 6 different sets of potential parameters, the corresponding values of the tensor to the scalar ratio ($r$) and the spectral index ($n_s$) are well inside the $2-\sigma$ bound of $Planck'18$.  In Fig. (\ref{plot3}), we have plotted $f_{PBH}$ with $M_{PBH}$ for three different values of $x$ and $m$ where, each set of potential parameter corresponds to a different color contour. In Fig. (\ref{fpbhvsmpbh-1}),   the red contour corresponds to the $x= 0.45018, m=3/5$ and green corresponds to $x= 0.614, m=1/2$ in both cases the $f_{PBH}$ is around $1 \%$, whereas the blue contour corresponds to $x= 0.50117, m=2/5$ and the corresponding $f_{PBH}$ is around $10 \%$. But that mass range is disfavoured to a certain extent by the HSC and OGLE experiment. Whereas in Fig (\ref{fpbhvsmpbh-2}), the red contour corresponds to $x=6$ and $m =-1/2$ which gives the negligible value of $f_{PBH}$ whereas the green contour for $x=8.4$ and $m=-1/2$ gives large value of $f_{PBH}$ around 50 $\%$ but again some part of it is disfavoured by the HSC. However, the blue contour for $x=8.6$ and $m=-3/5$ has some interesting results, where the $f_{PBH}$ is around $10 \%$ and it falls in the window of PBH mass where the $\Omega_{PBH}$ can be equal to $\Omega_{DM}$.

\underline{$q= +1$ case:}

In Fig(\ref{plot4}), we have plotted the dark matter fraction ($f_{PBH}$) against the mass of primordial black holes ($M_{PBH}$), three different colors correspond to the different combinations of potential parameters( $x,m$). In Fig (\ref{fpbhvsmpbh-3}) the purple contour corresponds to  $x= -2.14990,m=1/25$ which can give $100 \%$ of the present  dark matter density today. Again, some portion is disfavoured by the HSC and OGLE experiments, where  color cyan which stands for  $x= -1.69995,m=1/20$ gives the negligible value of $f_{PBH}$. However, the yellow contour shows the results for the combination $x=-1.1709,m=3/50$. In this case, the total $\Omega_{DM}$ today can be attributed to $\Omega_{PBH}$.  Whereas, in Fig (\ref{fpbhvsmpbh-4}) the purple contour corresponds to  $x= -1.1222,m =3/100$ gives about $0.1 \%$ of current dark matter density, the cyan contour stands for the  $x= -1.45,m=1/100$ which can give $f_{PBH}$ of around $1 \%$ but it has been bounded by the Extragalactic Gamma Background $(EG~\gamma-bkg)$ . Where for the combination $x=-1.5432,m=1/25$ the produced PBHs has the mass around $10^{-21}$ of \(\textup{M}_\odot\), which have been evaporated already through Hawking radiation. 

In this work, we have shown for a particular class of single-field inflationary model,  one can satisfy the condition that is necessary to produce sufficient PBHs in a vast mass range to the extent for a certain choice of class parameter, $\Omega_{PBH}$can be equal to the $100\%$ of total DM density today. Obviously, the details of reheating after the inflationary phase need to be studied and check if the model remains still viable or not or if it has any part to play in the case of producing the rest of the DM density through the mechanism proposed in \cite{koushik}. In our work, we have shown that the inflationary observables are well satisfied following the $Planck'18$ results. There are few avenues one can explore following this work. One can probe the parameter space in more detail, to check the theoretically viable form of the model to put it under the umbrella of string theory. One of the recent hooplas in theoretical physics, regarding swampland conjectures(SC) and trans-Planckian censorship conjecture(TCC) \cite{vafa1,vafa2,das1,kehagias,palma,dias} puts single field inflation directly in conflict. There are many solutions proposed in literature \cite{mrgy,das2,wali}. So one can check this class of model's consistency with SC and TCC, to claim the theoretical regularity. Secondly, as mentioned above, the reheating phase should be studied and one can follow \cite{skmrg} if one wants to construct SC and TCC consistent inflationary model in the domain of Randall Sundrum cosmology \cite{mathews,okada1,okada2}.  There is an interesting path to study in the case of PBHs production through the resonant particle creation and its implications on the CMB scale following \cite{9212,1701,0406}. After the proposal of tachyon inflation in \cite{sen1,sen2,sen3}, the noncanonical realization of different inflationary models have gained growing interest. One can follow \cite{0235}, to study the implication of noncanonical dynamics to populate the Universe with PBHs and their effects on the CMB observations through the sound speed($c_s$).

Thus, in a nutshell, the study of the production mechanism of PBHs have rich phenomenological implications as the scale of seeds of the PBH production is otherwise impossible to probe. We have defined a whole class of models with a large parameter space to explore. Keeping the observational demands of the Planck'18 mission in mind, we have narrowed down our search mostly to the parameter space where the inflationary observables are at least $2- \sigma$ consistent observationally. We have shown that this phenomenological single field model can get into the structure of a string inspired models of inflation for the right choice of parameters, but as we have said before, we refrain ourselves from studying any particular model, rather we have focused on showing this whole class can have potential to produce the correct ambience for PBHs' production. We have mentioned many possible avenues to explore after this. We will come back with such studies starting with the generalised reheating analysis very soon.
\section{Acknowledgement}
Work of MRG is supported by the Department of Science and Technology, Government of India under the Grant Agreement number IF18-PH-228 (INSPIRE Faculty Award) and by Science and Engineering Research Board(SERB), Department of Science and Technology(DST), Government of India under the Grant Agreement number CRG/2020/004347(Core Research Grant). The authors would like to thank Rahul Kumar, Imtiyaz Ahmad Bhat and Sukannya Bhattacharya for their useful discussion.

\end{document}